\documentclass[conference]{IEEEtran}
\IEEEoverridecommandlockouts

\usepackage{cite}
\usepackage{amsmath,amssymb,amsfonts}
\usepackage{algorithmic}
\usepackage{graphicx}
\usepackage{textcomp}
\usepackage{caption}
\usepackage{xcolor}
\usepackage{siunitx}
\usepackage{url}

\usepackage{cleveref}
\crefname{equation}{Eq.}{Eqs.}
\crefname{section}{Sec.}{Sec.}
\crefname{figure}{Fig.}{Figs.}
\crefrangeformat{equation}{Eqs. (#3#1#4)-(#5#2#6)}
\crefrangeformat{section}{Sec. (#3#1#4)-(#5#2#6)}
\crefrangeformat{figure}{Figs. #3#1#4-#5#2#6}

\usepackage[normalem]{ulem}

\def\BibTeX{{\rm B\kern-.05em{\sc i\kern-.025em b}\kern-.08em
    T\kern-.1667em\lower.7ex\hbox{E}\kern-.125emX}}
\begin{document}

\title{Stochastic Modeling of Power-Grid Frequency Fluctuations in Low-Inertia Systems via a Gaussian-Core Potential and Superstatistics}

\author{\IEEEauthorblockN{1\textsuperscript{st} Wanru Hao{*}}
\IEEEauthorblockA{\textit{Centre for Complex Systems} \\
\textit{Queen Mary University of London}\\
London, United Kingdom \\
w.hao@qmul.ac.uk}
\and
\IEEEauthorblockN{2\textsuperscript{nd} Alessandro Lonardi}
\IEEEauthorblockA{\textit{Centre for Complex Systems} \\
\textit{Queen Mary University of London}\\
London, United Kingdom \\
a.lonardi@qmul.ac.uk}
\and
\IEEEauthorblockN{3\textsuperscript{rd} Christian Beck}
\IEEEauthorblockA{\textit{Centre for Complex Systems} \\
\textit{Queen Mary University of London}\\
London, United Kingdom \\
c.beck@qmul.ac.uk}
}
\maketitle

\begin{abstract}

Power grid frequency stability is fundamental to the secure operation of modern energy systems, yet the growing penetration of renewables and the associated reduction of system inertia have made frequency fluctuations increasingly non-Gaussian and difficult to model. Existing stochastic models based on standard Ornstein--Uhlenbeck-type restoring terms yield a unimodal frequency distribution and therefore fail to reproduce the bimodal structure, central suppression, and heavy tails widely observed in empirical data. Here, we propose a data-driven stochastic process that combines a Gaussian-core potential with superstatistical modeling, assuming slowly fluctuating coefficients for the grid dynamics. The Gaussian-core potential captures the potential barrier that gives rise to the characteristic double-peak structure of frequency distributions. Fitting the model to frequency data resolved at one-second intervals from the Great Britain grid, we find that the central barrier parameter increases substantially from 2020 to 2025 as the grid inertia progressively decreases. To simulate superstatistics, we use an Euler--Maruyama discretization and sample the drift amplitude from a lognormal distribution, thereby successfully reproducing empirical bimodality and heavy tails, as well as the autocorrelation decay. Our results establish a compact and interpretable model for characterizing the evolving complexity of low-inertia grid frequency dynamics.

\end{abstract}



\section{Introduction}

Power grid frequency stability is a key prerequisite for the secure operation of energy systems. Stable grid operation requires a continuous balance between electricity supply and demand: excessive power raises the frequency at which the grid operates, whereas insufficient supply causes it to drop~\cite{kundur1994power}.

Power fluctuations arise from multiple sources, including transnational energy transactions in electricity markets \cite{mayer2018electricity} and industrial and residential consumption \cite{anvari2022data}. In recent years, the large-scale integration of renewable energy sources, such as wind and solar, has further intensified frequency fluctuations. Indeed, unlike conventional synchronous generators, such as thermal power plants, renewable generators are typically connected to the grid via electronic converters that lack rotational inertia. The loss of inertia weakens the grid's capacity to damp perturbations, thereby increasing the risk of frequency instability~\cite{ratnam2020future}.
In light of these evolving challenges, we are tasked with a better understanding of the nature of frequency fluctuations and with constructing accurate models of their dynamics and resulting statistical and temporal signatures.

Considerable progress has been made in this direction \cite{schaefer2018non,schafer2018isolating,vorobev2019deadbands,anvari2020stochastic,delgiudice2021effects,gorjao2020data,gorjao2020open,gorjao2021phase,gorjao2023stochastic,haehne2019propagation,kraljic2023towards,wen2025nonstandard,lonardi2026understanding,oberhofer2023non,oberhofer2025nonlinear}. In particular, heavy tails in frequency distributions, now widely recognized as a distinctive feature, have been modeled using superstatistical approaches \cite{beck2001dynamical,beck2003superstatistics} that account for fluctuating coefficients in the stochastic process governing frequency dynamics \cite{schaefer2018non}. Heavy tails also arise from the inclusion of fractional noise in stochastic dynamics \cite{kraljic2023towards}, from Lévy-stable distributions \cite{schaefer2018non}, or from simulations that incorporate a dynamically varying power imbalance between generation and load, which drives frequency fluctuations \cite{anvari2020stochastic,gorjao2020data}.

Multimodality of frequency distributions, also observed empirically, has been reproduced in simulations that include a time-varying power imbalance and nonlinear generator control \cite{vorobev2019deadbands,kraljic2023towards,oberhofer2023non}. In particular, these models assume that fluctuations are damped by a frequency-dependent function whose intensity increases either piecewise linearly \cite{vorobev2019deadbands,kraljic2023towards} or polynomially \cite{oberhofer2023non} with the magnitude of the fluctuations. Under similar assumptions, an analytical model predicting multimodal frequency distributions has recently been developed \cite{lonardi2026understanding}. In addition, frequency-dependent noise amplitude coefficients can also give rise to multimodal distributions \cite{oberhofer2025nonlinear}.

Assuming a time-varying power imbalance also affects the frequency autocorrelation, enabling the reproduction of recurrent jumps associated with energy transactions \cite{anvari2020stochastic,gorjao2020data}. Finally, power-law decay of the autocorrelation at long time lags has been modeled using fractional noise \cite{kraljic2023towards}.

Despite these efforts, open questions remain: can the coupling between control actions and power imbalance be captured by a single interpretable drift term in the stochastic differential equation modeling grid frequency? Furthermore, can such a formulation simultaneously reproduce multimodality, heavy tails, and the autocorrelation structure? This is particularly relevant for obtaining a compact description of grid frequency dynamics that reduces model complexity while capturing both stationary statistics (the distribution) and dynamical properties (the autocorrelation).

Here, we propose a data-driven stochastic model that incorporates a Gaussian-core potential \cite{stillinger1976phase}, which is commonly employed in condensed matter physics \cite{lang2000fluid,ruppeiner2021thermodynamic}. The potential enables us to accurately fit the annual frequency distributions of the Great Britain grid using six years of data. Interestingly, measured empirical data and their fitting distributions exhibit increasing bimodality over the years, which we interpret in the context of the progressive integration of renewables and associated reduction in the grid rotational inertia. We also interpret the potential shape in terms of the combined effects of time-varying power imbalance and nonlinear grid control, drawing on the literature \cite{wen2025nonstandard,lonardi2026understanding}.

To reproduce heavy tails and short-range autocorrelation decay, we use superstatistical modeling \cite{beck2001dynamical,beck2003superstatistics}. In particular, we simulate stochastic frequency dynamics by imposing a slowly fluctuating deterministic drive, given by the derivative of the potential. Fluctuating grid parameters reflect non-stationary operating conditions of generators arising, for instance, from time-varying consumer demand.

Taken together, our results provide a physically interpretable description of grid dynamics that accurately captures key statistical features of large-scale data measured at regimes with diminishing inertia. 


\section{Stochastic Model of Grid Frequency Dynamics}

Frequency dynamics is typically described by the Aggregated Swing Equation (ASE) \cite{ulbig2014impact}, which directly follows from energy conservation and governs the stochastic evolution of the bulk angular velocity of the grid $\omega(t)$. The angular velocity is related to the grid frequency $f(t)$ as $\omega(t) = 2\pi (f(t) - f_{\mathrm{R}})$, where the nominal frequency is $f_{\mathrm{R}} = \qty{50}{Hz}$ in UK. We write the ASE as
\begin{equation}
    \label{eq:swing_equation}
    \frac{d\omega}{dt} = H(\omega) + \epsilon \xi \,.
\end{equation}

In \Cref{eq:swing_equation}, $H(\omega)$ is the frequency-control term representing the restoring effect of generators that control frequency fluctuations, while $\xi \sim \mathcal{N}(0,1)$ denotes Gaussian white noise and $\epsilon > 0$ is the noise amplitude. The noise term captures stochastic fluctuations that are not resolved by control.

Under the model specified by the ASE, setting $H(\omega)$ as prescribed by nominal regulations \cite{ofgemnote} is not sufficient to reproduce heavy tails and multimodality \cite{lonardi2026understanding}. Therefore, we approach the problem from a different, data-driven perspective and infer $H(\omega)$ directly from measurements. In particular, we introduce the probability distribution $p(\omega, t)$, which follows a Fokker--Planck equation readily derived from the ASE under the assumption of Gaussian white noise \cite{risken1989fokker}:
\begin{equation}
    \label{eq:fp}
    \frac{\partial p}{\partial t}
    =
    -\frac{\partial}{\partial \omega}\left(H(\omega)p(\omega,t)\right)
    +
    \frac{\epsilon^2}{2}\frac{\partial^2 p}{\partial \omega^2} \,.
\end{equation}

During regular operation, the grid is approximately in a stationary state. Formally, this allows us to impose $\partial p / \partial t = 0$ in \Cref{eq:fp}. The stationary equation resulting from this condition can be integrated twice with respect to $\omega$ and yields the Gibbs--Boltzmann stationary distribution
\begin{equation}
    \label{eq:stationary_p}
    p(\omega)
    \propto
    \exp \left(
    \frac{2}{\epsilon^2}\int H(\omega)\,d\omega
    \right) \,.
\end{equation}

Conveniently, \Cref{eq:stationary_p} can be reformulated by introducing an effective potential $V(\omega)$, whose derivative corresponds to the negative of the deterministic control term $H(\omega)$:
\begin{equation}
    \label{eq: potential_definition}
    \frac{dV}{d\omega} = -H(\omega) \,.
\end{equation}
Defining $\beta = 2 / \epsilon^2$ and substituting \Cref{eq: potential_definition} into \Cref{eq:stationary_p} yields
\begin{equation}
    \label{eq:boltzmann_distribution}
    p (\omega) \propto \exp(-\beta V(\omega)) \,,
\end{equation}
which is the standard Gibbs--Boltzmann form with effective potential $V(\omega)$ and inverse temperature $\beta$.

Our goal is to find an analytically expressible potential $V(\omega)$ to fit experimental data and, then, to perform the fit with Maximum Likelihood Estimation (MLE) \cite{mlecode}.


\section{Gaussian-Core Potential Fitting}

The dataset used in this study consists of 1-second-resolution frequency time series from the Great Britain power grid, provided by NESO (National Energy System Operator)~\cite{data}. Great Britain is chosen as the case study because it has a large and electrically independent grid with substantial renewable-energy penetration, making it particularly relevant for investigating frequency fluctuations under low-inertia conditions. We use six years of data, from 2020 to 2025, totaling approximately 190 million data points, to validate the proposed model.

\begin{figure}[t]
    \centering
    \includegraphics{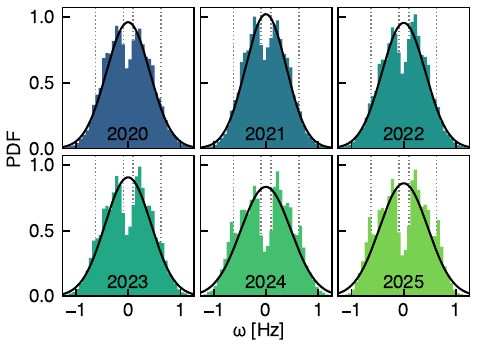}
    \caption{Empirical frequency histograms fitted with Gaussians (black lines). Dotted lines at $\pm 2\pi\cdot\qty{0.015}{Hz}$ and $\pm 2\pi\cdot\qty{0.1}{Hz}$ indicate the deadband and control-regime boundaries.}
    \label{fig:gaussian_fit_6years}
\end{figure}

\begin{figure}[t]
    \centering
    \includegraphics{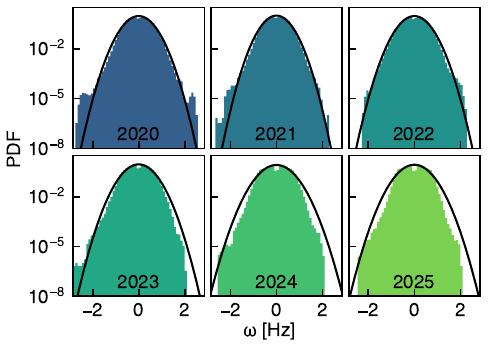}
    \caption{Same as in \Cref{fig:gaussian_fit_6years} but in a semi-logarithmic plot}.
    \label{fig:log_gaussian_fit_6years}
\end{figure}

Across all six years considered, the empirical frequency histograms exhibit noticeable non-Gaussian features, as illustrated in \Crefrange{fig:gaussian_fit_6years}{fig:log_gaussian_fit_6years}. In the earlier years, heavy tails are particularly pronounced. Their relative magnitude decreases as the frequency distribution gets progressively more multimodal in 2024 and 2025. At the same time, the central suppression of the distribution becomes more evident each year.

Secondary peaks, which become clearly distinguishable in 2025, arise because nominal control regulations in Great Britain prescribe two distinct control regimes \cite{ofgemnote}, the second of which is stronger and activated when frequency deviations exceed $\qty{0.1}{Hz}$ from the nominal value, as shown by \Cref{fig:gaussian_fit_6years}. The central suppression, instead, appears within the frequency deadband, a region of width $\qty{0.015}{Hz}$ where control is not activated due to finite-precision mechanisms of generators \cite{rebours2007survey}. In principle, it may not be immediately clear why the absence of control produces a dip in the frequency histogram. We address this point later in the discussion of our results.

The salient point of these initial remarks is that a Gaussian distribution is insufficient to fit frequency measurements.

Motivated by data, we employ a Gaussian-core potential model \cite{stillinger1976phase}:
    \begin{equation}
    \label{eq:gaussian_core_potential}
    V(\omega \mid \sigma, A)
    =
    \frac{\omega^2}{2\sigma^2}
    +
    A\exp\left[-\left(\frac{\omega}{\omega_0}\right)^2\right] \,.
\end{equation}
In \Cref{eq:gaussian_core_potential}, $\omega_0$ is the core width controlling the spread of the barrier around $\omega = 0$. We fix it using the nominal deadband size, namely, $\omega_0 = \qty{0.015}{Hz}$. We optimize $(\sigma, A)$ with MLE.

The two terms on the right-hand side of \Cref{eq:gaussian_core_potential} jointly confine the overall spread of the distribution through a quadratic potential (the first Gaussian term), while also introducing a central correction that accounts for the suppression observed in the frequency distribution (the second ``core'' term).

Consistent with nominal regulations, away from the core, when $|\omega| \gg \omega_0$, the potential prescribes linear control since $H(\omega) = - dV / d \omega \sim - \omega/\sigma^2$.

\begin{figure}[t]
    \centering
    \includegraphics{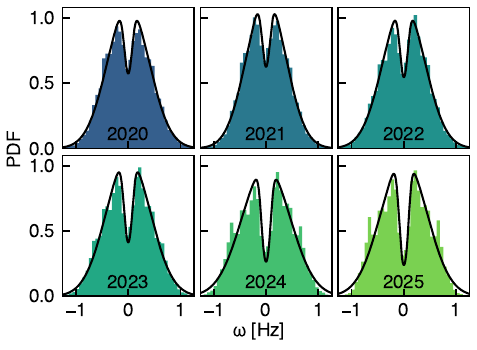}
    \caption{Empirical frequency histograms fitted with the Gaussian-core model of \Cref{eq:gaussian_core_potential}.}
    \label{fig:potential_fit_6years}
\end{figure}

\begin{figure}[t]
\centering
\begin{minipage}[t]{0.5\linewidth}
    \vspace{0pt}
    \centering
    \includegraphics{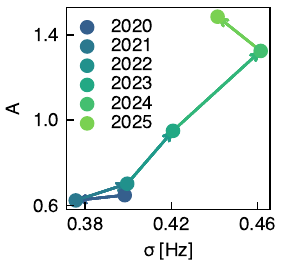}
    \caption{Estimated potential parameters across years in the $({\sigma}, {A})$ plane.}
    \label{fig:param_scatter}
\end{minipage}
\hfill
\begin{minipage}[t]{0.44\linewidth}
    \vspace{0pt}
    \centering
    \begin{tabular}{ccc}
    \hline
    Year & ${\sigma}$ $\unit{[Hz]}$ & ${A}$ \\
    \hline
    2020 & 0.3983 & 0.6490 \\
    2021 & 0.3755 & 0.6244 \\
    2022 & 0.3994 & 0.7025 \\
    2023 & 0.4207 & 0.9501 \\
    2024 & 0.4614 & 1.3239 \\
    2025 & 0.4414 & 1.4842 \\
    \hline
    \end{tabular}
    \captionof{table}{Estimated potential parameters truncated to arbitrary precision.}
    \label{tab:gaussian_core_params}
\end{minipage}
\end{figure}

In the fitting stage, we adopt the convention $\beta = 1$ (equivalently, we absorb $\beta$ into the parameters $(\sigma, A)$). Having fixed this last convention, we can perform MLE. We show the MLE fit for all years in \Cref{fig:potential_fit_6years} and their optimal parameters in \Cref{fig:param_scatter} and \Cref{tab:gaussian_core_params}.

Fig~\ref{fig:potential_fit_6years} shows that the Gaussian-core potential accurately describes the central suppression around the frequency deadband consistently for all years, improving upon the Gaussian fit of \Cref{fig:gaussian_fit_6years} or alternative unimodal fits used in the literature \cite{schaefer2018non}.

As reported in \Cref{tab:gaussian_core_params} and shown in \Cref{fig:param_scatter}, both fit parameters roughly monotonically increase from 2020 to 2025 for the NESO data \cite{data}. This quantifies the progressively larger spread and stronger central local minimum of the frequency distributions that move away from Gaussianity. 
Such an effect can be interpreted with superstatistical modeling that combines nonlinear control with a slowly varying power imbalance between generation and demand \cite{lonardi2026understanding} (as also suggested by Wen et al. \cite{wen2025nonstandard}).

The general idea behind superstatistical modeling is that many natural and engineered systems operate in non-stationary environments, where the relevant control parameters fluctuate on timescales that are long compared to their local relaxation dynamics. Such systems equilibrate locally, before slowly evolving parameters drive transitions between different quasi-stationary states. Such ideas have broad applicability across physics and engineering \cite{beck2001dynamical,beck2003superstatistics}.

In power grids, frequency can be described as locally equilibrating around quasi-stationary states associated with temporarily fixed power imbalances, while these imbalances evolve slowly over longer timescales and induce a drift. The resulting empirical distributions observed, for instance, in \Cref{fig:potential_fit_6years} emerge as a superposition of such local equilibria. Under these assumptions, the dip in the frequency histogram is analytically predicted to emerge within the deadband, where there is no control \cite{lonardi2026understanding}. 

\begin{figure}[t]
    \centering
    \includegraphics{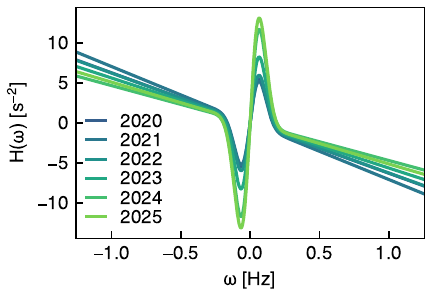}
    \caption{Control calculated as $H(\omega) = dV(\omega \mid \sigma, A) / d \omega$.}
    \label{fig:vprime}
\end{figure}

Importantly, this perspective also provides a physical interpretation for the Gaussian-core potential. Outside the deadband, the potential confines the grid to a quadratic well (equivalently, a linear restoring force), consistent with nominal linear control regulations. In the deadband, however, where control is absent, the potential becomes locally repulsive, capturing the combined influence of inactive regulation and slow power-imbalance fluctuations that drive the system away from $\omega = 0$. We show this effect by plotting the inferred control action $H(\omega)$ in \Cref{fig:vprime}. Notably, the repulsive core becomes particularly pronounced in later years, consistent with what we might expect in grids with greater renewable penetration, which leads to more volatile power imbalances and lower inertia. Simultaneously, the change of the effective potential over the years, in a sense, visualizes the energy transition quantitatively.

\section{Superstatistical Simulation}

While being accurate in the deadband, the Gaussian-core model does not reproduce heavy tails. We address this issue with superstatistical modeling. Additionally, we aim to reproduce the measured autocorrelation structure of the data in a physically meaningful and quantitatively reasonable way. Therefore, we extend the fitted Gaussian-core model by introducing a slowly varying superstatistical coefficient into the stochastic dynamics by following the approach of Schäfer et al. \cite{schaefer2018non}. In particular, we assume that both the potential parameters $(\sigma, A)$ vary slowly in time, reflecting the underlying non-stationary operating conditions of the grid.

We take 15 minutes as the characteristic timescale over which these parameters fluctuate, motivated by the typical resolution of energy market transactions. In fact, the Great Britain grid exchanges large amounts of energy at regular intervals, every 15, 30, and 60 minutes \cite{epexukmarket}, inducing large frequency fluctuations at the trading time points.
In practice, we discretize \Cref{eq:swing_equation} with the Euler--Maruyama scheme with a fixed integration time step $\Delta t = 1$ and resample a ``slow parameter'' $\alpha_n$ every $N = 15 \cdot 60$ steps (over a total of approximately $26\cdot10^7$ steps). We then rescale the coefficients of the fitted Gaussian-core potential by $\alpha_n$, thereby modulating both the confinement strength and the depth of the Gaussian-core suppression. In formulas, let $\omega_k$ denote the simulated frequency state at time step $k$, with time increment $\Delta t$. Then, $\omega_k$ is updated as
\begin{equation}
    \label{eq:em_simulation}
    \omega_{k+1}
    =
    \omega_k
    -
    \alpha_n\frac{\epsilon^2}{2} \frac{d V(\omega \mid \sigma, A)}{d\omega} \Bigg|_{\omega = \omega_k} \Delta t
    +
    \epsilon \sqrt{\Delta t} \,\xi_k \,
\end{equation}
where $\xi_k \sim \mathcal{N}(0,1)$ are i.i.d. random variables.

We draw $a_n$ from a lognormal distribution as in hydrodynamic turbulence models \cite{beck2007statistics}. This choice conveniently enforces parameter positivity, is standard in superstatistical modeling, and is also empirically consistent with measured fluctuations of generator drift coefficients \cite{schaefer2018non}.

We numerically choose the lognormal parameters to reproduce the empirical distributions and decay of the autocorrelation. In particular, we take $\alpha_n \sim \mathrm{LogNormal}(\mu =0, \sigma=0.35)$ (median $\exp(\mu) = 1$).

We note that, in the MLE fitting stage, we adopted the convention $\beta = 1$ in order to reconstruct the shape of $V(\omega)$ directly from the empirical distribution. For numerical simulations, however, this normalization is no longer imposed, and the coupling between the potential scale and the noise amplitude must be reintroduced explicitly. Formally, we fitted $V_{\mathrm{MLE}}(\omega \mid \sigma, A) = -\beta V(\omega)$ using MLE, where $V(\omega)$ is the true physical potential. We omitted the subscript $V_{\mathrm{MLE}}$ in \Cref{eq:gaussian_core_potential} for convenience. That is to say, the noise amplitude can now be reintroduced.

The stationary distribution associated with \Cref{eq:em_simulation} is independent of $\epsilon$, since $\epsilon$ cancels between drift and diffusion in its corresponding Fokker--Planck equation (\Cref{eq:fp}). However, $\epsilon$ controls the effective time scale of fluctuations and therefore affects the autocorrelation decay of the simulated signal. To match the empirical data, we set $\epsilon = 0.03$.

\begin{figure}[!t]
    \centering
    \includegraphics{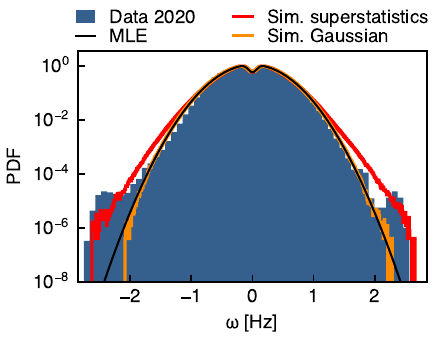}
    \caption{Comparison between empirical and simulated frequency distributions for 2020. Results from the superstatistical simulation are in red, those from the simulation with fixed parameters in orange, and the MLE fit from \Cref{fig:potential_fit_6years} is in black.}
    \label{fig:sim_pdf_6years}
\end{figure}

We first assess the simulation results in terms of the stationary distribution. \Cref{fig:sim_pdf_6years} shows the empirical frequency histogram of 2020 compared with the superstatistical simulation in \Cref{eq:em_simulation} and the MLE fit of \Cref{fig:potential_fit_6years}. We also run a simulation with fixed $\alpha_n = 1$ as a sanity check, namely, to verify that its simulated distribution matches the MLE prediction. We focus on 2020 since the empirical data display distinct heavy tails.

Overall, the superstatistical simulated distribution reproduces the main empirical features well. In particular, the model continues to capture the characteristic bimodal structure of the frequency distribution. In addition, we now observe heavy tails akin to those observed in the NESO measurements. Contrarily, the MLE fit and the simulation with constant $\alpha_n$ exhibit Gaussian tails and therefore underestimate the histograms in the extremes.
It should be noted that, in earlier years like 2020, the agreement between empirical and simulated distributions is particularly good as empirical distributions are roughly bimodal (see \Cref{fig:potential_fit_6years}). Such an agreement tends to be less robust in later years, particularly in 2024 and 2025, when multimodal distributions with additional structures emerge, and our predictions are valid only at a coarser level.

\begin{figure}[t]
    \centering
    \includegraphics{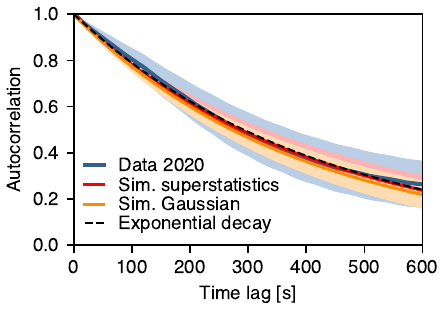}
    \caption{Comparison between empirical and simulated autocorrelations for 2020. Data are grouped into daily batches. For each day, we compute the autocorrelation and then average the results; the mean autocorrelation and its standard deviations are indicated in blue. The red and orange lines correspond to simulated autocorrelations, also averaged over batches of the same size as the empirical data. The black dashed line shows a fitted exponential decay $\exp(a \tau)$, with $\tau$ time lag.}
    \label{fig:sim_acf_6years}
\end{figure}

In addition to the stationary distribution, we evaluate the model using the frequency autocorrelation. The autocorrelation measures how strongly the frequency fluctuations remain correlated with their past values over different time lags, and therefore provides important information about the memory of the underlying stochastic processes.

\begin{figure}[t]
    \centering
    \includegraphics{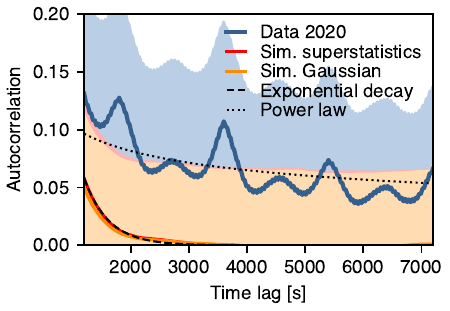}
    \caption{Data, simulation and exponential decay fit are as in \Cref{fig:sim_acf_6years}. The black dotted line is a power law fit $\tau^{-a}$, with $\tau$ time lag.}
    \label{fig:sim_acf_6years_long_range}
\end{figure}

\Cref{fig:sim_acf_6years} compares the empirical and simulated autocorrelations for the 2020 data at short time lags, namely below 10 minutes. Overall, our simulations reproduce the main decay pattern of the empirical correlations well. In particular, both simulations capture the well-known short-range exponential decay of frequency autocorrelation measures \cite{schaefer2018non,lonardi2026understanding,kraljic2023towards,gorjao2020open,gorjao2020data,anvari2020stochastic,wen2025nonstandard,oberhofer2023non}. The long-range autocorrelation patterns, shown in \Cref{fig:sim_acf_6years_long_range} between 20 minutes and 2 hours, are more complex. Here, recurrent energy transactions at 15-minute intervals induce an oscillatory structure in the autocorrelation function. At a coarse-grained level, the autocorrelation exhibits a slow decay consistent with a power law, which can be obtained by introducing fractional noise instead of Gaussian noise in \Cref{eq:swing_equation}, at the cost of reduced tractability and the loss of a closed-form Fokker--Planck equation as in \Cref{eq:fp} \cite{kraljic2023towards}. At present, our simulations underestimate these long-range patterns and do not capture the empirical autocorrelation.

\section{Conclusion and Discussion}

In this paper, we proposed a data-driven stochastic model for grid-frequency fluctuations in renewable-dominated, low-inertia power grids. The model combines a Gaussian-core effective potential with a superstatistical description of slowly varying grid parameters. The results show that this model can successfully reproduce the main empirical characteristics of the observed frequency distributions, including peaks, deadband suppression, heavy tails, and the short-range autocorrelation decay pattern, namely a rapid initial exponential decay.

Our findings are important because they demonstrate the effectiveness of a concise, physically interpretable model for stochastic frequency fluctuations in a real power system with progressively lower inertia, such as the Great Britain grid. The model performs well on a large-scale dataset spanning six years, comprising approximately 190 million data points, in which frequency fluctuations exhibit increasingly non-Gaussian behavior. Our results and considerations become increasingly important in light of the ongoing global transition toward higher shares of renewables, where frequency fluctuations are expected to become even more pronounced.

At the same time, the data also reveal features that are not yet fully captured by the present model. In recent years, measured frequency distributions have begun to exhibit four peaks at the boundary between nominal control regimes (\Cref{fig:gaussian_fit_6years}). In addition, the empirical autocorrelations show long-range recurrent oscillations (\Cref{fig:sim_acf_6years_long_range}). How to characterize these features using our model is an open question for future work. In particular, more complicated potential forms and explicit time-dependent market forcings \cite{gorjao2020data,kraljic2023towards,oberhofer2023non} can be considered.

\bibliographystyle{IEEEtran}
\bibliography{references}

\end{document}